\documentclass[aps,prl,twocolumn,superscriptaddress]{revtex4}
\usepackage{epsfig}
\usepackage[dvipsnames,usenames]{color}
\usepackage{color}
\usepackage{amsmath}
\usepackage{amssymb}
\usepackage{hyperref}
\usepackage{graphicx}

\usepackage{wasysym}
\usepackage{times}
\usepackage{comment}
\usepackage{array}
\usepackage{multirow}
\usepackage{tabularx}
\usepackage{float}
\usepackage[utf8]{inputenc}
\usepackage[T1]{fontenc}
\usepackage{cancel}
\usepackage{ulem}

\emergencystretch=\maxdimen
\begin{document}
\title{Hole clustering and mutual interplay in three-band Hubbard model}
\author{Mi Jiang}
\email[]{jiangmi@suda.edu.cn}
\affiliation{Institute of Theoretical and Applied Physics, Jiangsu Key Laboratory of Thin Films, School of Physical Science and Technology, Soochow University, Suzhou 215006, China}
\author{Yi-feng Yang}
\email[]{yifeng@iphy.ac.cn}
\affiliation{Beijing National Laboratory for Condensed Matter Physics and Institute of
Physics, Chinese Academy of Sciences, Beijing 100190, China}
\affiliation{University of Chinese Academy of Sciences, Beijing 100190, China}
\affiliation{Songshan Lake Materials Laboratory, Dongguan, Guangdong 523808, China}
\author{Guang-Ming Zhang}
\email[]{zhanggm@shanghaitech.edu.cn}
\affiliation{School of Physical Science and Technology, ShanghaiTech University, Shanghai 201210, China}
\affiliation{Department of Physics, Tsinghua University, Beijing 100084, China}

\begin{abstract}
Recent scanning tunnelling spectroscopy (STS) experiments revealed remarkable role of a supercell consisting $4\times4$ CuO$_2$ unit cells in the emergence of local nematic state and preformed local Cooper pairs and phase coherent cuprate superconductivity. By employing the numerically exact determinant Quantum Monte Carlo simulations, we mimic the effects of experimental Ca vacancy by an external local potential to investigate the charge and spectral properties of the system hosting two doped holes. The model numerically support the role of the $4\times4$ supercell as the building block of hole doped cuprates via the hole density distribution and local spectra around the local potential.
Our results might provide a theoretical support on the experimental observations and a platform for investigating local charge order and local Cooper pairs on the $4\times4$ supercell as the plausible route to understanding unconventional cuprate superconductivity.
\end{abstract}

\maketitle

{\bf Introduction.}
Despite of tremendous efforts in past three decades, the basic physics of high-$T_c$ cuprates remains a mystery. Experimentally, a number of exotic phenomenon including the pseudogap, the charge order, the strange metal, and the superconductivity, have been extensively explored but are still largely unexplained. In theory, most previous studies have been based on the one-band Hubbard model or the effective $t$-$J$ model, which can be derived by considering the Zhang-Rice singlet~\cite{ZhangRice} formed by  doped holes and Cu-3$d_{x^2-y^2}$ spins. While some insights have been drawn from these studies, there seems to be increasing debates concerning their validity in realistic materials under hole doping~\cite{nonZhangRice2017,T-CuO,Lau,Hadi1,Hadi2,Mi2020}. In particular, recent state-of-the-art calculations found that superconductivity may not be the ground state of the one-band Hubbard model on a square lattice~\cite{QinPRX}, at least in its simplest version. A smoking gun experiment is therefore crucial in order to identify the basic building block of the cuprate physics.

Recent scanning tunneling spectroscopy (STS) experiment has visualized an atomic-scale electronic state induced by hole dopants in Ca$_2$CuO$_2$Cl$_2$~\cite{Yayu1}, which implied that two introduced holes distribute around each Ca vacancy to form a $4a_0\times4a_0$ cluster with emergent low-energy spectral weight within the charge transfer gap. Upon increasing doping, these clusters grow in number and connect to form superconducting islands with checkerboard pattern even in the insulating sample. It was suggested that they may host local Cooper pairs whose spatial wave functions would overlap at large doping and eventually give rise to the phase coherent superconducting condensate~\cite{Yayu2}. This scenario of accounting for the cuprate superconductivity (SC) via preformed Cooper pairing has long been discussed and supported by a number of experimental probes~\cite{Tranquada2022}. In fact, the latest spin resolved tunnelling measurements on extremely underdoped Bi$_2$Sr$_{2-x}$La$_x$CuO$_{6-\delta}$ provided further support on this type of physical picture of an electronic structure with $4a_0\times4a_0$ basic plaquette as local pairing platform~\cite{HaihuWen}.

\begin{figure}[t!]
\psfig{figure=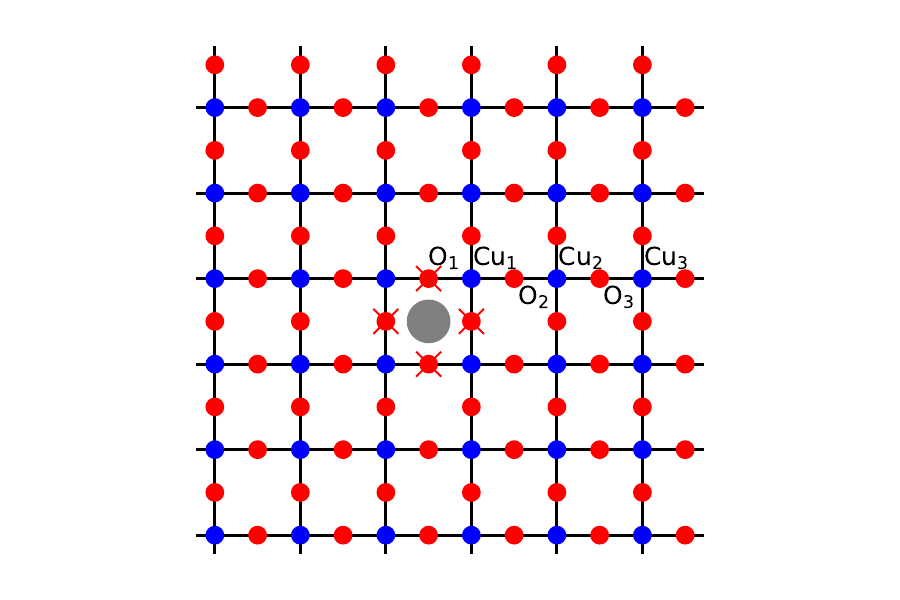,height=6cm,width=.46\textwidth, trim={2.0cm 0.5cm 1.0cm 0}, clip} 
\caption{Lattice geometry with the central plaquette hosting the local potential $V$ labeled as red crosses. Inequivalent ``ring'' of Cu or O sites are labeled by O$_m$ and Cu$_m$ with $m=1,2,3$ etc.}
\label{fig1}
\end{figure}

These experimental progress has motivated a few theoretical studies, for example, on the effects of impurity on the one-hole doped in-gap states in the single-band Hubbard model~\cite{Lijianxin}. In particular, it was found that a repulsive impurity potential can move the in-gap state from the lower Hubbard band towards the upper Hubbard band upon increasing the potential strength, which is qualitatively consistent with the experimental finding that the in-gap states are close to upper Hubbard band instead of expected lower Hubbard band. Nonetheless, this previous work only focussed on the spectral properties of one-hole doped system. Besides, the discussion of the most fascinating experimental feature of $4a_0\times4a_0$ plaquette formation is absent.

In this work, we investigate the implication of the STS experiment in the more realistic theoretical framework based on three-band $d$-$p$ model together with local potential, which is proved to be an effective model of accounting for the local physics observed in experiments. In particular, we point out the importance of local impurity potentials mimicking the Ca vacancy and explain successfully the observed cluster structure as the building block of hole-doped cuprates. Our calculations confirm the formation of $4a_0\times4a_0$ basic plaquette and reveal a possible percolation picture for understanding the basic physics of the high-$T_c$ cuprates.

{\bf Model and methodology.}
We adopt the celebrated three-band Hubbard model~\cite{Emery,tpd2016} with additional local potential consisting of Cu-3$d_{x^2-y^2}$ and O-2$p_{x/y}$ orbitals in hole language
\begin{eqnarray}
    {\cal H} &=& \sum\limits_{\langle ij \rangle \sigma} t_{ij}
(d^{\dagger}_{i\sigma}p_{j\sigma}^{\vphantom{dagger}}
+ p^{\dagger}_{j\sigma}d_{i\sigma}^{\vphantom{dagger}}) \nonumber \\
&+& \sum\limits_{\langle jj' \rangle \sigma} t_{jj'}
(p^{\dagger}_{j\sigma}p_{j'\sigma}^{\vphantom{dagger}}
+ p^{\dagger}_{j'\sigma}p_{j\sigma}^{\vphantom{dagger}}) \nonumber \\
&+& U_{dd} \sum\limits_{i} n^{d}_{i\uparrow} n^{d}_{i\downarrow} + (\epsilon_d-\mu) \sum\limits_{i\sigma} n^{d}_{i\sigma} \nonumber \\
&+& U_{pp} \sum\limits_{i} n^{p}_{i\uparrow} n^{p}_{i\downarrow} + (\epsilon_p-\mu) \sum\limits_{i\sigma} n^{p}_{i\sigma} 
- V \sum\limits_{i\in A,\sigma} n^{p}_{i\sigma} 
\label{model}
\end{eqnarray}
where $d^{\dagger}_{i\sigma}(d_{i\sigma}^{\vphantom{dagger}})$
and $p^{\dagger}_{i\sigma}(p_{i\sigma}^{\vphantom{dagger}})$ 
are hole creation (destruction) operators for Cu-3$d_{x^2-y^2}$ and O-2$p_{x/y}$ orbitals in unit cell $i$ with spin $\sigma$, and
$n^{d,p}_{i\sigma}$ are the associated number operators. The chemical potential $\mu$ can be tuned for a desired average occupancy. In  hole language, the hopping phase convention is
\begin{eqnarray}
t_{ij} &=& t_{pd} (-1)^{\eta_{ij}}
 \nonumber \\
t_{jj'} &=& t_{pp} (-1)^{\beta_{jj'}}
\label{}
\end{eqnarray}
where $\eta_{ij}=1$ for $j=i+\frac{1}{2} \hat{x}$ or $j=i-\frac{1}{2} \hat{y}$ and $\eta_{ij}=0$ for $j=i-\frac{1}{2} \hat{x}$ or $j=i+\frac{1}{2} \hat{y}$. Additionally, $\beta_{jj'}=1$ for $j=i+\frac{1}{2} \hat{x}+\frac{1}{2} \hat{y}$ or $j=i-\frac{1}{2} \hat{x}-\frac{1}{2} \hat{y}$ and $\beta_{jj'}=0$ for $j=i-\frac{1}{2} \hat{x}+\frac{1}{2} \hat{y}$ or $j=i+\frac{1}{2} \hat{x}-\frac{1}{2} \hat{y}$. We emphasize that all physical quantities in this work will be presented in hole language so that the undoped system corresponds to $n_h=1$. $U_{dd}$ and $U_{pp}$ denote the local repulsive interaction of Cu-3$d_{x^2-y^2}$ and O-2$p_{x/y}$ orbitals respectively while we adopt $U_{pp}=0$ for simplicity. The site energies $\epsilon_d$ and $\epsilon_p$ control the relative occupancy of $d$ and $p$ orbitals to be consistent with the realistic cuprates. 

For clarity, Fig.~\ref{fig1} depicts the lattice geometry with the local potential $V$ included in the central plaquette. To respect the C$_4$ rotational symmetry, we label inequivalent ``ring'' of Cu or O sites by O$_m$ and Cu$_m$ with $m=1,2,3$ and so on. 
To account for the experimental findings of the hole dopants in Ca$_2$CuO$_2$Cl$_2$, it is important to notice that Ca cation is located on top of the plaquette center of the Lieb lattice of CuO$_2$ units and its vacancy introduces an effective potential on surrounding Cu and O ions (gray circle in Fig.~\ref{fig1})~\cite{Yayu1,Yayu2}. For simplicity, we restrict this potential only to the four nearest surrounding oxygen sites, which are labeled as $A$ in Eq.~\ref{model} and denoted as cross symbols in Fig.~\ref{fig1}. In this way, a finite $V$ would attract more holes to dope onto neighboring sites of the Ca vacancy, namely four oxygen sites in the central plaquette. 

To fully take into account all the energy scales on the equal footing, we adopt the principally exact numerical technique of finite temperature determinant Quantum Monte Carlo (DQMC)~\cite{blankenbecler81} to explore the physics of Eq.~\ref{model}, particularly the essential role of local potential $V$ in shaping the electronic properties of cuprates. As a well established and widely used computational method, DQMC provides an approximation-free solution of a number of challenging problems due to the presence of strong electronic correlations. We will see that this local potential is crucial for explaining the observed local electronic structure pattern in experiment, indicating that underdoped cuprates are inhomogeneous on a fundamental level beyond the clean models. Throughout the paper, we focus on a characteristic parameter set for the cuprates (in units of eV): $U_{dd}=8.0$, $U_{pp}=0.0$, $t_{pd}=1.3$, $t_{pp}=0.49$, $\epsilon_d=0.0$, and $\epsilon_p=3.24$ at inverse temperature $\beta=10$. A more realistic parameters sets do not qualitatively modify our major findings but may induce more severe QMC sign problem~\cite{tpd2016}. All presented data were performed on a $N=8 \times 8$ lattice with periodic boundary. Since the local charge distribution is limited largely in a $4\times 4$ cluster, the finite-size effects should not be significant.
Distinct from the previous investigation studying the case of one-hole doping~\cite{Lijianxin}, we focus on the system with two doped holes to investigate if their electronic structure naturally form a $4\times 4$ cluster to coincide with the STS experiments. 


\begin{figure}[h!]
\psfig{figure=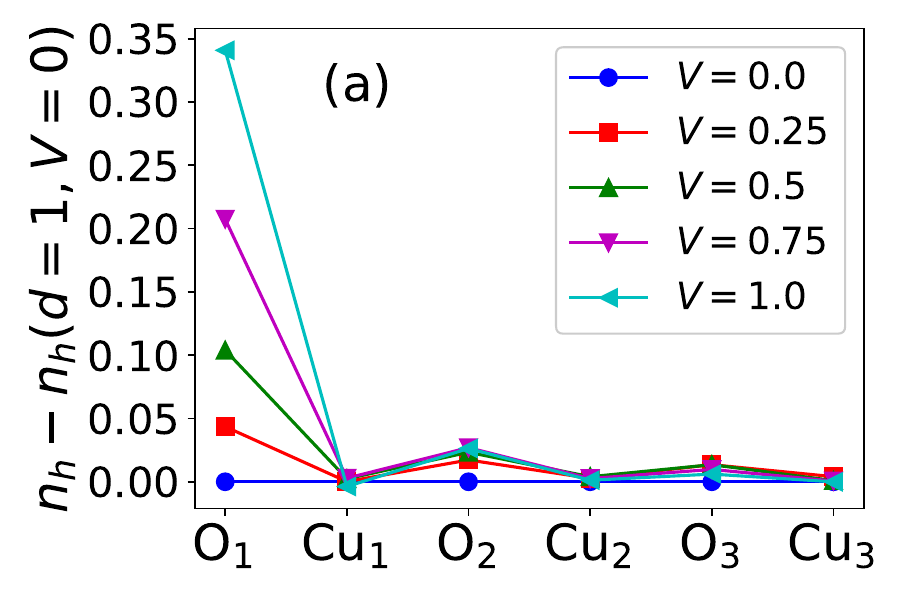,height=3.0cm,width=.235\textwidth, trim={0.0cm 0 0 0}, clip} 
\psfig{figure=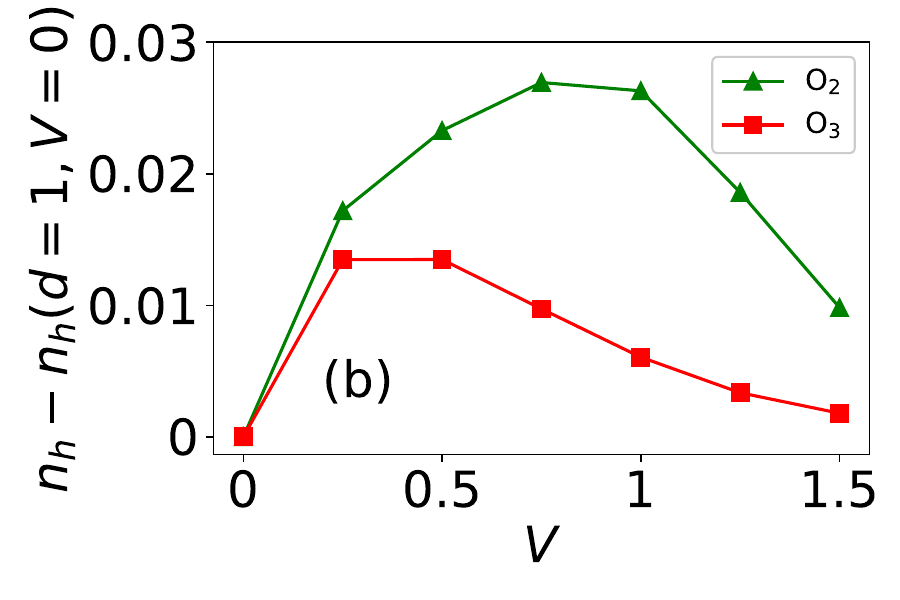,height=3.0cm,width=.235\textwidth, trim={0.0cm 0 0 0}, clip} 
\psfig{figure=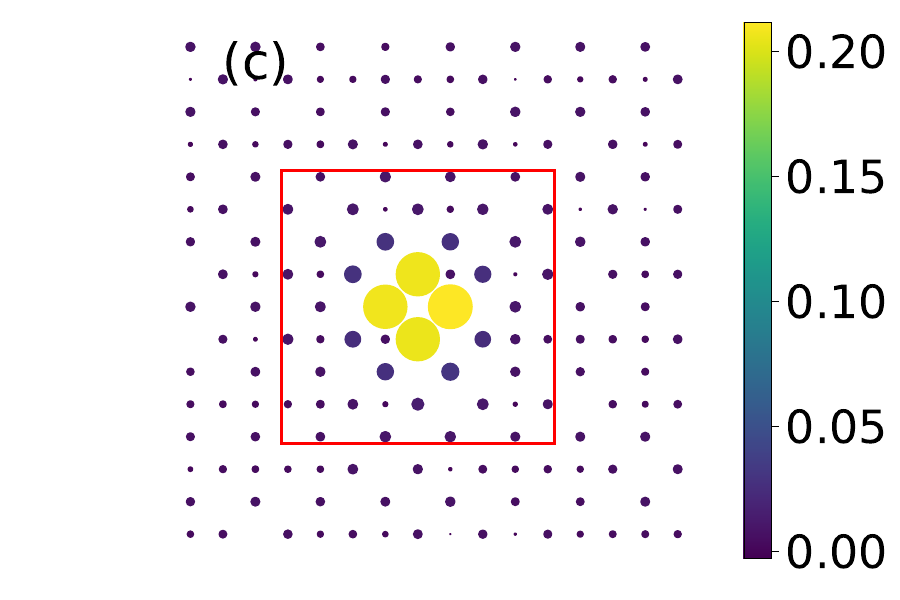,height=3.4cm,width=.235\textwidth, trim={2.5cm 0 0 0}, clip} 
\psfig{figure=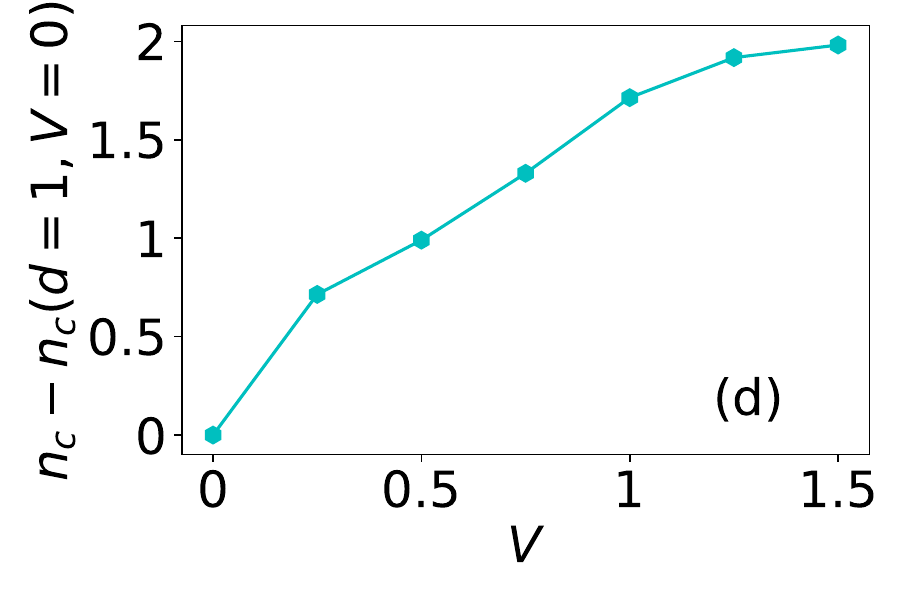,height=3.0cm,width=.235\textwidth, trim={0.0cm 0 0 0}, clip} 
\caption{(a) Evolution of the charge density of two doped holes subtracted by the undoped system without local potential at various inequivalent locations (see Fig.~\ref{fig1}); (b) Hole density of O$_2$ and O$_3$ sites with varying $V$; (c) Spatial distribution of hole charge density at the characteristic $V=0.75$ with red square labeling the $4\times 4$ cluster; (d) Total doped hole density within the $4\times 4$ cluster (red square) with increasing $V$.}
\label{fig2}
\end{figure}

{\bf Hole density distribution.}
To explore the impact of local potential $V$, we first examine the hole density distribution. Fig.~\ref{fig2}(a) compares the local hole charge density (averaged over all equivalent Cu and/or O sites) of the system hosting {\bf two} doped holes (from a single Ca vacancy) for different local potential after subtracting those of the undoped system without local potential. For $V=0$, the charge distribution is homogenous on all Cu or O sites. Applying the local potential $V$ attracts the doped holes preferentially on the innermost O$_1$ rings. For large enough $V$, the doped holes are almost solely trapped on O$_1$, in which case some of the original holes on Cu$_1$ (due to hybridization) even transfer to O$_1$. For moderate $V$, we find that the hole density of O is neither homogeneous nor trapped on O$_1$, but decays with the distance from the central plaquette and shows finite distribution also on O$_2$ and O$_3$ rings. By contrast, the hole density on Cu sites is hardly affected because of the strong correlation $U_{dd}$.

A better view on the evolution of the hole distribution may be seen in Fig.~\ref{fig2}(b). A non-monotonic dependence of the hole density change is seen on O$_2$ and O$_3$ with increasing local potential, separating the diagram roughly into three characteristic regions corresponding to small, moderate, and large $V$. Evidently, a moderate local potential can produce the density modulation such that it decays until O$_3$ sites, namely within a $4\times 4$ supercell around the central plaquette applied with local potential, to match with STS experimental findings. To see this more clearly, Figs.~\ref{fig2}(c) depicts the spatial distribution of $n_h$ subtracted by the undoped system. As shown by the red square, at typical $V=0.75$, the doped hole density decays and extends to a supercell consisting approximately $4\times 4$ d-p$_x$-p$_y$ orbital unit cell. Panel (d) further shows the evolution of the doped hole density within the supercell (red square in panel (c)) versus $V$. This comprises one of our major findings to account for the recent STS experiments, namely the building blocks (each containing two holes) that eventually form the phase coherent SC condensate are checkerboard puddles with a size around $4 a_0 \times 4 a_0$.

{\bf Local spectra.} To further understand the impact of local potential, we examine the spectral properties via the single-particle orbital-dependent local density of states (DOS) $A(\omega)$ relating to the local imaginary-time ordered Green's function
$G(\tau)=\sum\limits_{{\bf j}} \langle
\mathcal{T} c^{\phantom{\dagger}}_{{\bf j}}(\tau) c^{\dagger}_{{\bf
j}}(0) \rangle$ with $c$ denoting $d$ or $p$ orbital's hole via
\begin{equation}
   G(\tau)= \int_{-\infty}^{\infty}\ d\omega \frac{e^{-\omega\tau}}{e^{-\beta\omega}+1}\ A(\omega). \label{aw}
\end{equation}
The analytic continuation from $G(\tau)$ to $A(\omega)$ is performed via maximum entropy method \cite{gubernatis91}. Note that our spectral function is also obtained in hole language to be consistent with other quantities, which is distinct from the conventional electron language adopted in photoemission experiments.

\begin{figure} 
\psfig{figure=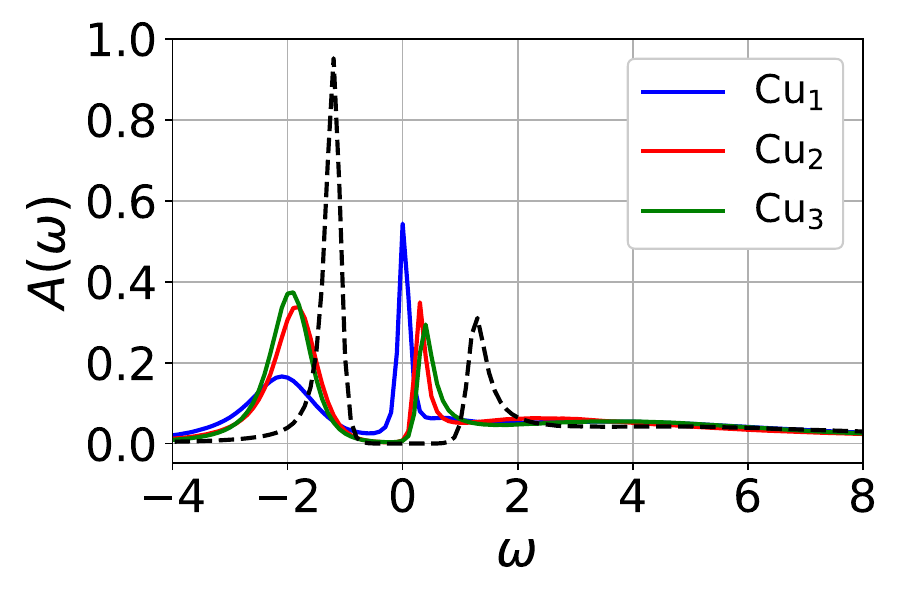,height=4.5cm,width=.45\textwidth, trim={0 1.4cm 0 0}, clip} 
\psfig{figure=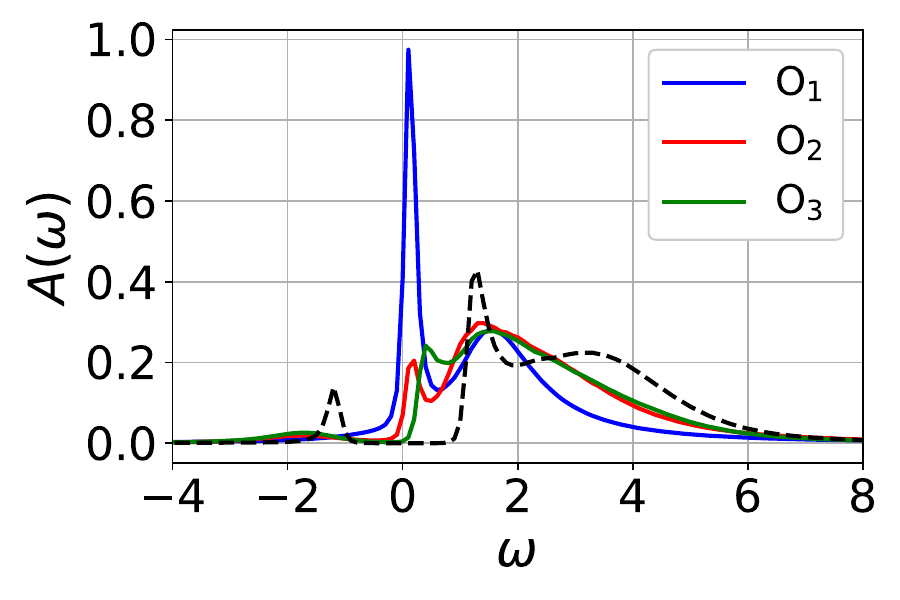,height=5.0cm,width=.45\textwidth, clip} 
\caption{Local density of states (DOS) of inequivalent Cu and O sites at $V=0.75$. The dashed lines denote the spectra of undoped system without local potential.}
\label{fig3}
\end{figure}

For the undoped system, all rings labeled as Cu$_m$ and O$_m$ in Fig.~\ref{fig1} are equivalent. As shown as the dashed lines of Fig.~\ref{fig3}, the spectral functions on Cu and O sites exhibit a charge transfer gap. The $d$-$p$ hybridization induces a broadening in the charge transfer band around $\omega\sim 1.2\,$eV from O-2$p$'s spectra, which is also reflected in the Cu-3$d$ orbital's spectral behavior at corresponding energy scales. However, once the local potential is included, low-energy spectral weights in $A(\omega)$ are seen to emerge within the charge transfer gap on all inequivalent Cu and O sites~\cite{Lijianxin}. A dominant feature is the spectral peak of O$_1$ and Cu$_1$ at the Fermi level, which might be associated with the Zhang-Rice singlet formed between doped holes on O$_1$ sites and their neighboring Cu$_1$ spins. Away from Ca vacancy (gray circle in Fig.~\ref{fig1}), the local DOS's peak shifts towards higher energy with tiny spectral weight at Fermi energy still on O$_2$ sites. The spectral features, particularly the peaks at Fermi energy, provide another indicator on the mobility of the two doped holes mainly residing on the sites nearest to the local potential in forming the superconducting states.  

\begin{figure}[t] 
\psfig{figure=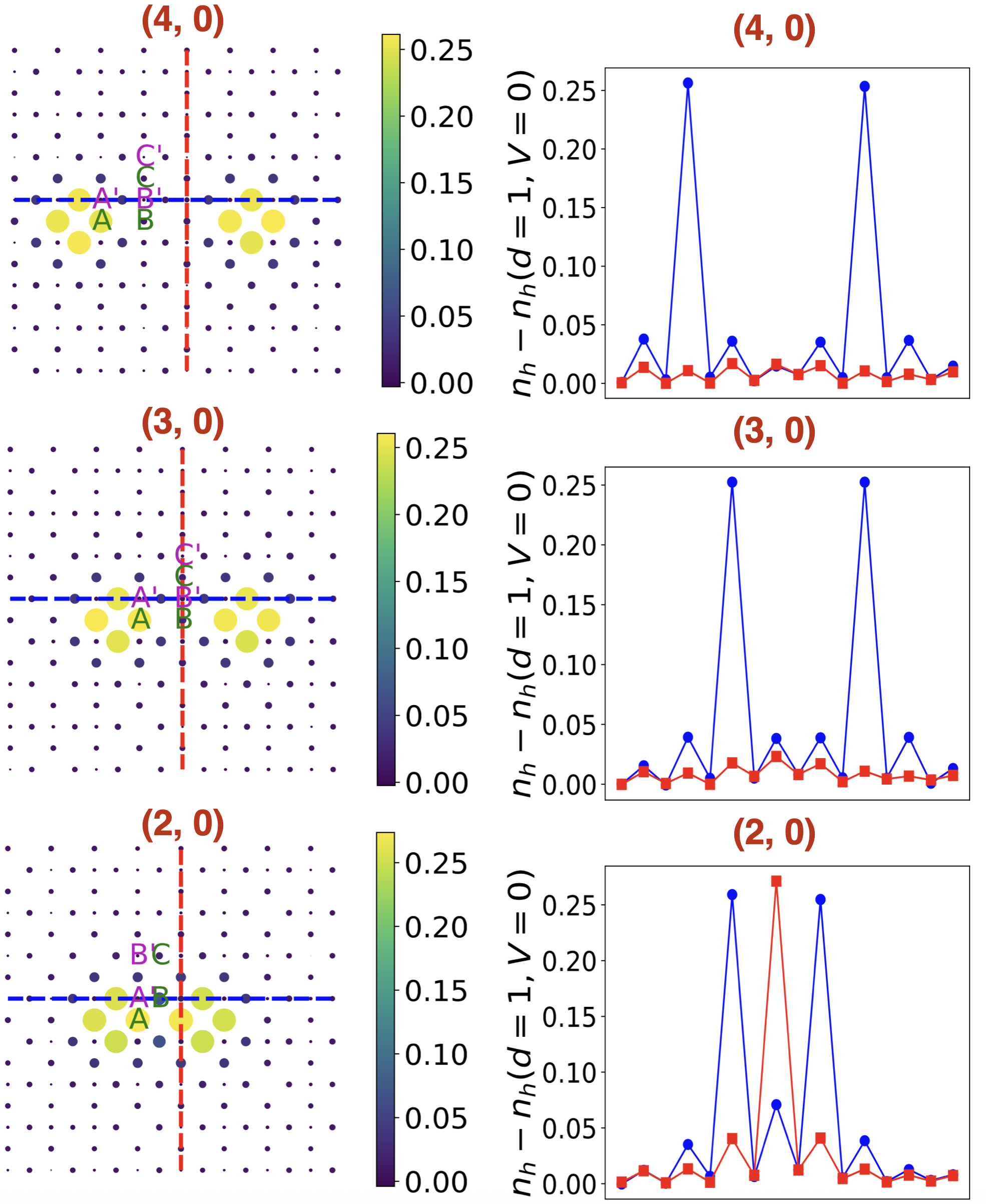,height=9.4cm,width=.48\textwidth, trim={2.5cm 0 0 0}, clip} 
\caption{Left: Hole density distribution (subtracted by that of undoped $V=0$ system) for two clusters of $(4,0), (3,0), (2,0)$ configurations from top to bottom.
Right: NN pair hopping along dashed lines labeled on the left panels. The local potential is fixed to $V=0.75$.}
\label{fig4}
\end{figure}

\begin{figure}[t] 
\psfig{figure=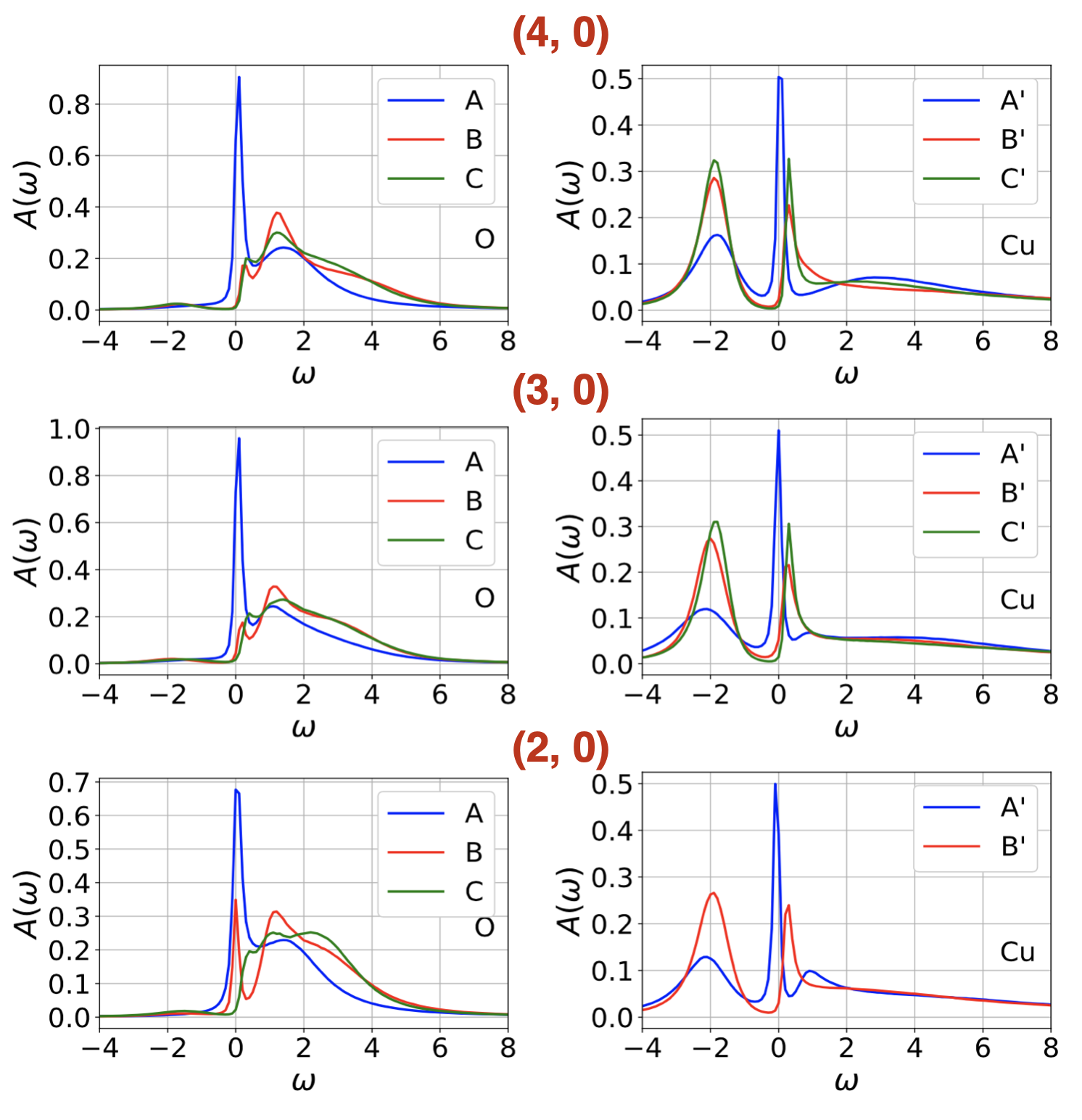,height=9.0cm,width=.48\textwidth, clip}
\caption{Local spectra for two clusters of $(4,0), (3,0), (2,0)$ configurations from top to bottom for typical Cu and O sites labeled in Fig.~\ref{fig4}'s left panels. The local potential is fixed to $V=0.75$.}
\label{fig5}
\end{figure}

{\bf Hole cluster pair.} 
We have demonstrated the capbility of two doped holes in forming a supercell consisting of $4\times 4$ d-p$_x$-p$_y$ orbital unit cells.
It can be naturally expected that the number of supercells increases upon more hole doping. As proposed by the STS experiment, the interplay between supercells plays a vital role for the development of both charge order and superconductivity. In particular, the supercell forms the building block for the charge order at 1/8 doping and may connect to form superconducting islands even in the insulating sample. 
To understand how this works, we explore the local electronic states by adding two local potentials of the same magnitude on different locations to explore their interaction with reducing distance. As shown in Fig.~\ref{fig4}, we choose three typical situations of a pair of Ca dopants~\cite{Yayu1}, with separating vector $(4,0), (3,0), (2,0)$ from top to bottom, respectively. 
The left panels illustrate the charge density distribution (subtracted by undoped system without $V$) of four doped holes; while the right panels show the evolution of hole density along specific red and blue dashed lines marked in the left panels. For $(4,0)$ configuration (top panel), the two clusters are almost independent as evidenced by the even distribution around local potentials. Correspondingly, the two repeated peaks in the right panel show up (blue line) at each local potential. Besides, tiny density distributes along the intermediate perpendicular direction (red line). As two clusters approach each other, e.g., for $(3,0)$ configuration in the middle row, the hole density accumulates gradually between two clusters indicating the mutual overlap to induce density redistribution. This trend is more clearly shown in the bottom panels, which provide the ultimate fate of two touched clusters with $(2,0)$ separation, which reveals the strong enhancement of hole charge density around the touching point of two neighboring clusters. With the arrangement of two clusters along a particular direction, the hole density naturally form a stripe-like distribution, which might provide some hints on the role played by the local potentials in the recent discovery of the local nematic states within a $4\times 4$ supercell~\cite{HaihuWen}.

Similar to the case of single cluster, Fig.~\ref{fig5} demonstrates the effects of the distance between two clusters via local spectral properties. For $(4,0)$ in top panel, the local spectra are all reminiscent of the single cluster case. For example, the spectra at A and A' points are almost the same as those at O$_1$ and Cu$_1$ points in Fig.~\ref{fig3}, reflecting the independence between two clusters. The approaching between two clusters has negligible impact on the local spectra for $(3,0)$ separation (middle panel), which is associated with the fact that the doped hole mainly resides on the O sites applied with local potential and their nearest Cu sites. Surprisingly, when two clusters touch with each other with $(2,0)$ separation (bottom panel), the dominant spectral peaks at Fermi energy of A and A' remain so that the hole density redistribution has seemingly negligible effects in spectral features.

{\bf Discussion and conclusion.} 
Due to the limitation of QMC simulations, we cannot go to the larger lattice size, doping regime, and much lower temperature to study the possible percolation of clusters and more importantly the superconducting pairing properties. However, our work can already provide some insight on this novel perspective of cuprate physics. It suggests that one may start from the $4\times4$ supercell with two holes and construct an effective low-energy model to describe the 1/8 charge order and high $T_c$ superconductivity. A most crude version of such model would be an effective $t$-$V$ model of bosons including a hopping term between different clusters and a local repulsive interaction to prevent bosons to collapse. Since the $4\times4$ supercell of two holes is exactly 1/8 doping, the long-range charge order would be naturally realized at 1/8 hole doping where the bosons may distribute uniformly on the lattice due to the ``local" repulsive interaction. In this sense, the 1/8 charge order may be viewed as an effective parent state of the superconductivity, away from which would introduce additional bosons to disrupt the charge order and cause superconducting condensate.

To summarize, we have investigated the hole density distribution and local spectral properties using the numerically exact DQMC simulations on a three-band $d$-$p$ model to understand the recent observation in STS experiment on hole-doped cuprates. We find it is important to introduce local impurity potential to mimic the Ca vacancy for hole doping. For an intermediate potential, our calculations can well reproduce the experimentally observed $4\times4$ supercell structure with its emergent low-energy spectral weight. The interplay between two supercells may lead to hole density redistribution. This suggests that the $4\times4$ cluster could be a basic building block of hole doped cuprates. It may naturally give rise to the long-range charge order at 1/8 hole doping. Our work provides an alternative way to view the cuprate physics and may serve as a novel platform for understanding the mystery of high $T_c$ cuprates. It is expected that the hole clustering would affect the pairing properties and ultimately global coherent superconductivity through percolation of clusters, the possibility of local pair formation relevant to recent experimental reports~\cite{Yayu2,HaihuWen} within our model is worthwhile exploring in detail.

{\bf Acknowledgments} 
This work was supported by the National Natural Science Foundation of China (12174278, 12474136), the National Key Research and Development Program of MOST of China (2017YFA0302902), the startup fund from Soochow University, the Priority Academic Program Development (PAPD) of Jiangsu Higher Education Institutions, and the Strategic Priority Research Program of CAS (Grant No. XDB33010100).



\begin{thebibliography}{27}
\expandafter\ifx\csname natexlab\endcsname\relax\def\natexlab#1{#1}\fi
\expandafter\ifx\csname bibnamefont\endcsname\relax
  \def\bibnamefont#1{#1}\fi
\expandafter\ifx\csname bibfnamefont\endcsname\relax
  \def\bibfnamefont#1{#1}\fi
\expandafter\ifx\csname citenamefont\endcsname\relax
  \def\citenamefont#1{#1}\fi
\expandafter\ifx\csname url\endcsname\relax
  \def\url#1{\texttt{#1}}\fi
\expandafter\ifx\csname urlprefix\endcsname\relax\def\urlprefix{URL }\fi
\providecommand{\bibinfo}[2]{#2}
\providecommand{\eprint}[2][]{\url{#2}}

\makeatletter
\providecommand \@ifxundefined [1]{%
 \@ifx{#1\undefined}
}%
\providecommand \@ifnum [1]{%
 \ifnum #1\expandafter \@firstoftwo
 \else \expandafter \@secondoftwo
 \fi
}%
\providecommand \@ifx [1]{%
 \ifx #1\expandafter \@firstoftwo
 \else \expandafter \@secondoftwo
 \fi
}%
\providecommand \natexlab [1]{#1}%
\providecommand \enquote  [1]{``#1''}%
\providecommand \bibnamefont  [1]{#1}%
\providecommand \bibfnamefont [1]{#1}%
\providecommand \citenamefont [1]{#1}%
\providecommand \href@noop [0]{\@secondoftwo}%
\providecommand \href [0]{\begingroup \@sanitize@url \@href}%
\providecommand \@href[1]{\@@startlink{#1}\@@href}%
\providecommand \@@href[1]{\endgroup#1\@@endlink}%
\providecommand \@sanitize@url [0]{\catcode `\\12\catcode `\$12\catcode
  `\&12\catcode `\#12\catcode `\^12\catcode `\_12\catcode `\%12\relax}%
\providecommand \@@startlink[1]{}%
\providecommand \@@endlink[0]{}%
\providecommand \url  [0]{\begingroup\@sanitize@url \@url }%
\providecommand \@url [1]{\endgroup\@href {#1}{\urlprefix }}%
\providecommand \urlprefix  [0]{URL }%
\providecommand \Eprint [0]{\href }%
\providecommand \doibase [0]{https://doi.org/}%
\providecommand \selectlanguage [0]{\@gobble}%
\providecommand \bibinfo  [0]{\@secondoftwo}%
\providecommand \bibfield  [0]{\@secondoftwo}%
\providecommand \translation [1]{[#1]}%
\providecommand \BibitemOpen [0]{}%
\providecommand \bibitemStop [0]{}%
\providecommand \bibitemNoStop [0]{.\EOS\space}%
\providecommand \EOS [0]{\spacefactor3000\relax}%
\providecommand \BibitemShut  [1]{\csname bibitem#1\endcsname}%
\let\auto@bib@innerbib\@empty
\bibitem{ZhangRice}
F. C. Zhang and T. M. Rice, Phys. Rev. B {\bf 37}, 3759 (1988).

\bibitem{nonZhangRice2017}
I. Santoso, W. Ku, T. Shirakawa, G. Neuber, X. Yin, M. Enoki, M. Fujita, R. Liang, T. Venkatesan, G.A. Sawatzky, A. Kotlov, S. Yunoki, M. Rübhausen, and A. Rusydi, Phys. Rev. B {\bf 95}, 165108 (2017).

\bibitem{T-CuO}
C.P.J. Adolphs, S. Moser, G.A. Sawatzky, and M. Berciu, Phys. Rev. Lett. {\bf 116}, 087002 (2016).

\bibitem{Lau}
B. Lau, M. Berciu, and G.A. Sawatzky, Phys. Rev. Lett. {\bf 106}, 036401 (2011).

\bibitem{Hadi1} 
H. Ebrahimnejad, G.A. Sawatzky, and M. Berciu, Nat. Phys. {\bf 10}, 951 (2014).

\bibitem{Hadi2} 
H. Ebrahimnejad, G. A. Sawatzky and M. Berciu, J. Phys.: Cond. Mat. {\bf 28}, 105603 (2016).

\bibitem{Mi2020}
M. Jiang, M. Moeller, M. Berciu and G. A. Sawatzky, Phys. Rev. B 101, 035151 (2020).

\bibitem{QinPRX}
M. Qin, C.-M. Chung, H. Shi, E. Vitali, C. Hubig, U. Schollwöck, S. R. White, and S. Zhang, Phys. Rev. X 10, 031016 (2020).

\bibitem{Yayu1}
H. Li, S. Ye, J. Zhao, C. Jin, and Y. Wang, Science Bulletin, 66, 1395 (2021).

\bibitem{Yayu2}
S. Ye, C.Zou, H. Yan, Y. Ji, M. Xu, Z. Dong, Y. Chen, X.J. Zhou, and Y. Wang, to be published (2022). 

\bibitem{Tranquada2022}
Y. Li, A. Sapkota, P. M. Lozano, Z. Du, H. Li et al, arXiv: 2205.01702 (2022).

\bibitem{HaihuWen}
H. Li, H. Li, Z. Wang, S. Wan, H. Yang, H.-H. Wen, arXiv: 2207.00783 (2022).

\bibitem{Lijianxin}
C.-P. He, S.-L. Yu, T. Xiang, and J.-X. Li, Chin. Phys. Lett. {\bf 39}, 057401 (2022).

\bibitem{Emery}
V.J. Emery, Phys. Rev. Lett. {\bf 58}, 2794 (1987).

\bibitem{tpd2016}
Y.F. Kung, C.-C. Chen, Y. Wang, E. W. Huang, E. A. Nowadnick et al, Phys. Rev. B 93, 155166 (2016).

\bibitem{blankenbecler81}
R. Blankenbecler, D.J. Scalapino, and R.L. Sugar, Phys. Rev. D24,
2278 (1981).

\bibitem{wenjian}
W. Hu, R. T. Scalettar, E. W. Huang, and B. Moritz, Phys. Rev. B 95, 235122 (2017).

\bibitem{gubernatis91}
J.E. Gubernatis, M. Jarrell, R.N. Silver, and D.S. Sivia, Phys. Rev. B 44, 6011 (1991).


\end{thebibliography}
\end{document}